\documentclass[onecolumn,12pt,journal,draftclsnofoot]{IEEEtran}
\usepackage[latin9]{inputenc}
\usepackage{color}
\usepackage{textcomp}
\usepackage{amsmath}
\usepackage{amssymb}
\usepackage{graphicx}
\PassOptionsToPackage{normalem}{ulem}
\usepackage{ulem}
\usepackage{colortbl}
\usepackage{stfloats}
\definecolor{mygray}{rgb}{.83, .83, .83}
\definecolor{mygreen}{rgb}{0.9608,    0.9608 ,   0.8627}
\definecolor{mycyan}{rgb}{0.9412,    1.0000,    1.0000}
\usepackage{setspace}

\makeatletter

\usepackage{amsfonts}
\usepackage{mathrsfs}
\usepackage{mathrsfs}\usepackage[noblocks]{authblk}
\usepackage[compress]{cite}
\usepackage{bm}
\usepackage{algorithm}
\usepackage{algorithmic}
\usepackage{epsfig}\usepackage{graphics}\usepackage{subfigure}\usepackage{epsfig}\usepackage{epstopdf}
\usepackage{graphics}\usepackage{subfigure}\usepackage{upgreek}\usepackage{caption}\usepackage{theorem}
\usepackage{breqn}
\usepackage{multicol}
\usepackage{eufrak}
\usepackage{eucal}
\usepackage{array}
\usepackage{cases}
\usepackage[compress]{cite}
\usepackage{algorithm}
\usepackage{algorithmic}
\usepackage{empheq}
\allowdisplaybreaks[4]
\theoremheaderfont{\normalfont\bfseries}

\makeatother

\begin{document}

\title{Channel Estimation for mmWave Pinching-Antenna Systems}
 
\author{Gui Zhou$^{\mathdollar}$, Vasilis Papanikolaou$^{\mathdollar}$, Zhiguo Ding$^\natural$, and Robert Schober$^{\mathdollar}$ \\
	
		\IEEEauthorblockA{
		$^{\mathdollar}$Friedrich-Alexander-Universit\"{a}t Erlangen-N\"{u}rnberg, Germany \\
		$^\natural$  University of Manchester, UK\\
		e-mail: \{gui.zhou,  vasilis.papanikolaou, robert.schober\}@fau.de, zhiguo.ding@manchester.ac.uk
	}
 \vspace{-9mm}
}

\maketitle
\begin{abstract}
The full potential of pinching-antenna systems (PAS) can be unblocked if pinching antennas can be accurately activated at positions tailored for the serving users', which means that   acquiring accurate channel state information (CSI) at arbitrary positions along the waveguide is essential for the precise placement of antennas. In this work, we propose an innovative channel estimation scheme for millimeter-wave (mmWave) PAS. The proposed approach requires activating only a small number of pinching antennas, thereby limiting antenna switching and pilot overhead. Specifically, a base station (BS) equipped with a waveguide selectively activates subarrays located near and far from the feed point, each comprising  a small number of pinching antennas. This configuration effectively emulates a large-aperture array, enabling high-accuracy estimation of multipath propagation parameters, including angles, delays, and path gains.  Simulation results demonstrate that the proposed method achieves accurate CSI estimation  and data rates while effectively reducing hardware switching and pilot overhead.
\end{abstract}


	 \vspace{-4mm}
\section{Introduction }

Pinching-antenna technology has emerged as a promising solution for building large-scale reconfigurable arrays with minimal hardware cost. Inspired by DOCOMO's 2022 demonstration \cite{decomo}, where beam patterns were formed by placing simple dielectric pinching elements (e.g., plastic clips) along a waveguide, the pinching antenna system (PAS) concept \cite{decomo,Ding-1} enables flexible beam control by dynamically adjusting the array configurations, including the number and the positions of the antennas. This low-cost, flexibly deployable design has sparked growing research interest in harnessing waveguide-based arrays for adaptive and efficient wireless communications. 

The authors of \cite{Ding-1}  provided the first  analytical framework to characterize the channel properties  and path-loss mitigation capabilities of pinching antennas, further demonstrating their potential for supporting non-orthogonal multiple access (NOMA), multiple-input multiple-output (MIMO), and interference management in multi-waveguide scenarios. In \cite{Ding-2}, discrete placement of pinching antennas was considered, where a NOMA-assisted downlink PAS scenario was considered and a matching-theory-based algorithm was developed to optimize the number and positions of activated antennas for throughput maximization. Building on this foundation, \cite{Ali} extended the investigation to multiuser MIMO settings, proposing joint hybrid beamforming and pinching-element placement strategies for both uplink and downlink, showing substantial throughput gains over conventional MIMO baselines.

Most existing studies on PAS primarily focused on communication performance analysis and optimization, and assumed the availability of perfect accurate channel state information (CSI). In PAS, a pinching antenna is expected to continuously vary its position within a designated area, implying that CSI at any arbitrary position, or across a dense set of predefined ports, must be known to fully exploit its potential. Consequently, conventional channel estimation methods designed for fixed-antenna systems become inapplicable due to the resulting high pilot and hardware overhead. Similarly, channel estimation techniques developed for fluid antenna systems (FAS) cannot be adopted, as they are tailored for a far-field channel model that depends only on angles of departure/arrival and path gains \cite{FASCE}. In contrast, PAS typically operate in the near-field due to the high frequency and the extended size of the waveguide, where users often lie within the Fresnel region; thus, a distance-dependent free-space channel model has to be adopted.  To date, limited research has addressed CSI estimation for PAS. In \cite{channel-1}, the user location was estimated and used for beam training, which is applicable to pure line-of-sight (LoS) only. In \cite{channel-2}, a machine learning-based method was proposed to predict the CSI at all waveguide locations, but  can incur substantial pilot overhead due to the $10^5$ training samples required for training dataset construction. These challenges underscore the urgent need for low-cost, scalable channel estimation schemes tailored to the unique characteristics of PAS.

In this work, we address the channel estimation problem for a time-division duplexing (TDD)  mmWave PAS, in which the base station (BS) is equipped with a waveguide to serve a user.  
Leveraging the sparse-scattering nature of mmWave channels, we propose a sparse array-assisted multi-path channel reconstruction method,  which involves  four steps.  First, by activating a few pinching antennas near the feed point, referred to  as the near-end subarray, the BS receives pilot signals from the user to estimate the corresponding angles of arrival (AoAs). Second, by activating  a few pinching antennas placed far from the feed point, forming the far-end subarray, the BS collects another set of pilot signals, from which the path distances are estimated based on  the previously acquired AoAs. Third, the path gains are estimated by combining the signals received from both subarrays. Finally, utilizing the free-space steering vector model and the estimated sparse parameters, the full channel vector between all potential antenna positions on the waveguide and the user is reconstructed. Simulation results demonstrate that the proposed method provides   highly accurate full CSI reconstruction, especially under strong LoS conditions (i.e., high Rician factor), while requiring low antenna switching and pilot overhead only.

\textbf{Notations:} The following mathematical notations and symbols
are used throughout this paper. Vectors and matrices are denoted by
boldface lowercase letters and boldface uppercase letters, respectively.
$\mathbf{X}^{*}$, $\mathbf{X}^{\mathrm{T}}$, and $\mathbf{X}^{\mathrm{H}}$ denote the conjugate, transpose, and Hermitian (conjugate transpose) of matrix $\mathbf{X}$, respectively. $||\mathbf{x}||_{2}$ denotes the L2-norm of vector $\mathbf{x}$. $\mathrm{Diag}(\mathbf{x})$ is a diagonal matrix with the entries
of $\mathbf{x}$ on its main diagonal.  The Hadamard product between
two matrices $\mathbf{X}$ and $\mathbf{Y}$ is denoted
by $\mathbf{X}\odot\mathbf{Y}$.  $\mathbb{C}$ denotes the complex field, $\mathbb{R}$ denotes the real field, and $j\triangleq\sqrt{-1}$ is the imaginary unit.  The notation $\mathcal{CN}(a,b)$ 
 refers to a circularly symmetric complex Gaussian distribution with mean $a$ and variance $b$.

	 \vspace{-3mm}
\section{System Model }

\begin{figure}
	\centering \includegraphics[width=3.4in,height=1.4in]{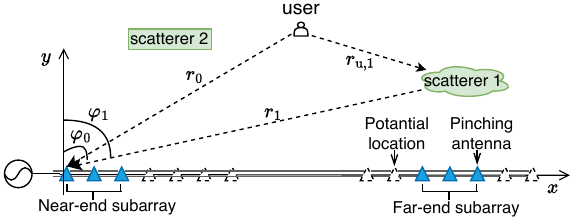}
	\captionsetup{font={small}} \caption{System model for channel estimation.}
	\label{system model} 
	 \vspace{-6mm}
\end{figure}

In this work, we consider  a PAS assisted mmWave communication
system  where the BS is equipped with a dielectric waveguide of length $L$
Radiation elements are created by locally deforming the waveguide, i.e., applying
pinching elements at one or multiple points, to induce controlled electromagnetic
leakage for signal radiation, and are therefore termed pinching antennas.
Let the radiation coefficient of the $m$th pinching antenna be denoted
by $w_{m}=e^{-j\beta x_{m}}$, where $x_{m}$ represents the distance
from the feed point. Here, $\beta=\eta_{g}\frac{2\pi}{\lambda}$ denotes
the propagation constant of the waveguide, where $\eta_{g}$
is the waveguide refractive index and $\lambda$ is the free-space wavelength. 

The configuration principle of the pinching array is to determine
the optimal number and the optimal spatial locations of the pinching antennas
on the waveguide, according to the predefined quality-of-service (QoS) metric,
such as spectral efficiency, energy efficiency, or
total power consumption. To enable optimal pinching array
configuration, it is crucial that the CSI
between any candidate position on the waveguide and the user is 
available at the BS, which motivates the design of channel estimation methods for PAS.

	 \vspace{-3mm}
\subsection{Channel Model}

Let $M$ denote the number of candidate positions of the pinching elements on the waveguide.  In theory, $M$ could be arbitrarily large, since pinching elements may be positioned anywhere along the waveguide. However, for practical deployment and to ensure feasibility, the elements must be discretized at half-wavelength intervals, limiting $M$ to $M = M = \tfrac{2L}{\lambda}$.
%In theory, $M$ is infinite, as pinching elements may be placed at arbitrary positions along the waveguide. However, for practical implementation, the activation positions are discretized with half-wavelength spacing, yielding a finite set of $M=2L/\lambda$ candidate points. 

Assuming a short-range, high-frequency application, the wireless channel is modeled by a free-space sparse multiple path model, which includes a LoS path and $P$ scattered paths{\footnote{For simplicity, it is assumed that the same scatterers are visible for the whole waveguide.}}.  Denote $h_{m}$ by the channel coefficent between the $m$th candidate  position and the user \cite{NF-channel-2}, where 
{\small
\begin{equation}
h_{m}\!=\frac{\lambda}{4\pi r_{0,m}}e^{-j\frac{2\pi}{\lambda}r_{0,m}}+\frac{\lambda}{(4\pi)^{3/2}}\!\sum_{p=1}^{P}\!\frac{\alpha_{p}}{r_{p,m}r_{{\rm u},p}}e^{-j\frac{2\pi}{\lambda}(r_{p,m}+r_{{\rm u},p})},\label{eq:ji}
\end{equation}}with  $\alpha_{p}>0$ being the radar cross section (RCS) of scatterer $p$.
 $r_{0,m}$ and
$r_{p,m}$ represent the physical distances from the user and the
$p$th scatterer, respectively, to the $m$th candidate position on
the waveguide.  $r_{{\rm u},p}$ is the physical distance from the user to the
$p$th scatterer. Without loss of generality, assume that the waveguide is placed along the $x$-axis
of the $xy$-plane, with its feed point located at the origin. For
this setup, the distance from the $p$th scatterer or the user to the $m$-th
candidate position is given by $r_{p,m}=\sqrt{r_{p}-2r_{p}x_{m}\sin\varphi_{p}+x_{m}^{2}}$
for $0\leq p\leq P$, where $\varphi_{p}$ denotes the AoA at a reference point and $r_{p}$ represents the distance
to the reference point. Correspondingly, the distance $r_{{\rm u},p}$ is calculated as  $r_{{\rm u},p}=\sqrt{r_0^2 - 2{r_0}{r_p}\cos(\varphi_{0}-\varphi_{p}) +r_p^2}$, for $1\leq p\leq P$. Without loss of generality, we designate the
first candidate location, which also corresponds to the feed point,
as the reference point, as shown in Fig. 	\ref{system model}. Thus, we have $r_{p}=r_{p,1}$, for $0\leq p\leq P$.

	 \vspace{-2.5mm}
\subsection{LS Channel Estimation}

To acquire the full CSI, denoted by $\mathbf{h}=[h_{1},...,h_{M}]^{\mathrm{T}}$,
a straightforward approach is to apply the least squares (LS) method given the received signals at all candidate locations. Specifically,  the user transmits pilot signal $s(t)$ over $M$ time slots, during
which the BS sequentially activates a pinching antenna at each of the $M$ candidate positions for signal reception. By collecting the
received signals over the $M$ time slots, the signal observed at the
BS can be expressed as \vspace{-1mm}
\begin{align}
\mathbf{y} & =[y(1),y(2),...,y(M)]^{\rm T}=\mathbf{w}\odot\mathbf{h}\odot\mathbf{s}+\mathbf{z}\nonumber \\
 & =\mathrm{diag}(\mathbf{w}\odot\mathbf{s})\mathbf{h}+\mathbf{z},\label{eq:jie}
\end{align}
where $\mathbf{w}=[w_{1},...,w_{M}]^{\mathrm{T}}$,   $\mathbf{s}=[s(1),...,s(M)]^{\mathrm{T}}$, constrained by  transmit power $q = ||\mathbf{s}||^2_2$,
and $\mathbf{z}$ denotes the additive white Gaussian noise (AWGN) vector with independent identically distributed (i.i.d.) entries following 
 $\mathcal{CN}(0,\sigma^{2})$ with noise power $\sigma^{2}$. 
Assuming that all pilot signals are
non-zero, the LS estimate of $\mathbf{h}$ is given by\vspace{-1mm}
\begin{align} 
\hat{\mathbf{h}}_{\rm LS} & =\mathrm{diag}(\mathbf{w}\odot\mathbf{s})^{-1}\mathbf{y}.\label{eq:u90}
\end{align}\vspace{-1mm}

For simple point-to-point signal models such as (\ref{eq:jie}), the
LS estimator in (\ref{eq:u90}) achieves high estimation accuracy,
approaching the Cramér--Rao Bound (CRB) in the high signal-to-noise ratio
(SNR) regime. However, this
comes at the cost of substantial pilot overhead, which scales linearly
with the number of candidate locations $M$. 

\subsection{Geometric Channel Structure}

To address this issue,  we exploit the geometric structure of the channel vector $\mathbf{h}$
based on (\ref{eq:ji}),  which can be rewritten as   \vspace{-1mm}
\begin{align*}
\mathbf{h} & =\sum_{p=0}^{P}\beta_{p}\mathbf{a}(r_{p},\varphi_{p}),
\end{align*}
where the complex path gains  are given by $\beta_{0}=\frac{\lambda}{4\pi r_{0}}e^{-j\frac{2\pi}{\lambda}r_{0}}$
and $\beta_{p}=\frac{\lambda\alpha_{p}}{(4\pi)^{3/2}r_{p}r_{{\rm u},p}}e^{-j\frac{2\pi}{\lambda}(r_{p}+r_{{\rm u},p})}$,
 $1\leq p\leq P$. Here, $\mathbf{a}(r_{p},\varphi_{p})\in\mathbb{C}^{M}$
denotes the general near-field array response vector \cite{NF-channel},  given by\vspace{-1mm}
\begin{equation}
\mathbf{a}(r_{p},\varphi_{p})\!=\!\!\left[\!1,\frac{r_{p}}{r_{p,2}}e^{-j\frac{2\pi}{\lambda}(r_{p,2}-r_{p})},\!\cdots\!,\frac{r_{p}}{r_{p,M}}e^{-j\frac{2\pi}{\lambda}(r_{p,M}-r_{p})}\right]^{\mathrm{T}}\!\!,\label{eq:n9j9}
\end{equation}
which depends on the user location for $p=0$ and  the scatterer positions
for $1\leq p\leq P$. The CSI for each path can be fully characterized
by estimating the parameter triplet $\{\beta_{p},r_{p},\varphi_{p}\}$.
In other words, estimating $3P$ parameters suffices, eliminating the need to estimate all $M$ individual channel entries. This leads to a classical sparse parametric estimation problem.

Motivated by near-field sparse channel estimation, the angles and
distances of each path can be estimated based on the near-field steering
vector (\ref{eq:n9j9}) using existing techniques such as codebook-based
search \cite{NFchannel} or subspace methods like MUSIC \cite{NFMUSIC}. To use (\ref{eq:n9j9}),
however, requires at least $N$ pinching antennas uniformly distributed along the waveguide with half-wavelength spacing, such that the aperture $D$ of the activated pinching array satisfies the Fresnel condition
$d=\frac{2D^{2}}{\lambda}\geq\max\{r_{p}\}_{p=1}^{P}$. 
For example, at 28\,GHz and a maximum target distance of 50\,m, the
required critical number of pinching antennas is $N=1+\sqrt{\frac{8r}{\lambda}}\approx195$.
This number is still substantial and, in practice, an even larger  $N$ is needed to ensure estimation accuracy. Therefore, it becomes  essential to develop channel estimation algorithms that not only reduce  the required pilot overhead and  the required number of activated pinching antennas, but also ensure reliable estimation accuracy.

	 \vspace{-2mm}
\section{Proposed Estimation Method }

In this section, inspired by sparse array design, we propose a scheme
that activates a small number of sparsely distributed pinching antennas
from the candidate locations to construct a large effective
aperture. The selected antennas are divided into two subarrays: a
near-end subarray with $M_{1}$ elements placed close to the feed
point for angle estimation, and a far-end subarray with $M_{2}$ elements
positioned farther away for distance estimation assisted by the estimated angles, as illustrated in Fig. 	\ref{system model}. The path gains are then recovered by jointly processing the signals from both subarrays.

For the near-end subarray, the first $M_{1}$ candidate locations  are selected  and the corresponding  pinching antennas are activated one by one. The BS sequentially  receives $M_{1}$ pilot signals from these antennas, such that the received
signal is given by\vspace{-1mm}
\begin{align}
\mathbf{y}_{\mathrm{ne}} & =\mathrm{diag}(\mathbf{w}_{\mathrm{ne}}\odot\mathbf{s}_{\mathrm{ne}})\mathbf{h}_{\mathrm{ne}}+\mathbf{z}_{\mathrm{ne}},\label{eq:u87}
\end{align}
where $\mathbf{h}_{\mathrm{ne}}$ and $\mathbf{w}_{\mathrm{ne}}$
are the channel and radiation vectors associated with the near-end
subarray, corresponding to the first $M_{1}$ entries of $\mathbf{h}$
and $\mathbf{w}$, respectively. $\mathbf{s}_{\mathrm{ne}}=[s(1),...,s(M_{1})]^{\mathrm{T}}$
contains the pilot signals, and $\mathbf{z}_{\mathrm{ne}}$ is the
AWGN vector at the near-end subarray.

Subsequently, the BS activates pinching antennas at candidate locations indexed from $m_{\mathrm{fe}}$ to $m_{\mathrm{fe}}+M_{2}-1$,
where $m_{\mathrm{fe}}$ is positioned near the far end of the waveguide,
receiving the pilot signal $\mathbf{s}_{\mathrm{fe}}=[s(M_{1}+1),...,s(M_{1}+M_{2})]^{\mathrm{T}}$.
The received signal is given by\vspace{-1mm}
\begin{align}
\mathbf{y}_{\mathrm{fe}} & =\mathrm{diag}(\mathbf{w}_{\mathrm{fe}}\odot\mathbf{s}_{\mathrm{fe}})\mathbf{h}_{\mathrm{fe}}+\mathbf{z}_{\mathrm{fe}},\label{eq:u89}
\end{align}
where $\mathbf{h}_{\mathrm{fe}}$ and $\mathbf{w}_{\mathrm{fe}}$
denote the channel and radiation vectors associated with the far-end
subarray, corresponding to entries $m_{\mathrm{fe}}$ through $m_{\mathrm{fe}}+M_{2}-1$
of $\mathbf{h}$ and  $\mathbf{w}$, respectively. $\mathbf{z}_{\mathrm{fe}}$
represents the corresponding AWGN vector. 
The total power of the pilot signal is constrained by $q = ||\mathbf{s}_{\mathrm{ne}}||^2_2+||\mathbf{s}_{\mathrm{fe}}||^2_2$.

	 \vspace{-2mm}
\subsection{Coarse Estimation of Angles and Distances}
\label{eq:n98999}
We begin by providing a higher-order approximation of the path distance  \cite{NF-channel}: 
\begin{equation}
r_{p,m}\approx r_{p}-x_{m}\sin\varphi+\frac{x_{m}^{2}\cos^{2}\varphi}{2r_{p}}+\mathcal{O}(x_{m}^{3}).\label{eq:ji6=00005C}
\end{equation}

\textit{Angle Estimation:} The near-end subarray is used to accomplish angle estimation by an approximation based on the far-field steering channel model, as explained in the following. Since $x_{m}$
is small compared to distance $r_p$, the higher-order terms $\mathcal{O}(x_{m}^{2})=\frac{x_{m}^{2}\cos^{2}\varphi}{2r_{p}}$ and $\mathcal{O}(x_{m}^{3})$
are negligible. Thus, the distance can be modelled by the first-order
Taylor approximation, as $r_{p,m}\approx r_{p,m}^{\mathrm{first}}=r_{p}-x_{m}\sin\varphi_{p}$,
$1\leq m\leq M_{1}$. Additionally, the free-space path loss ratio
$\frac{r_{p}}{r_{p,M_{1}}}\approx1$ is identical for the first $M_{1}$
antennas, due to their spatial proximity. Under these approximations,
the steering vector for the near-end subarray can be approximated
by a distance-independent far-field steering vector, given by
\begin{equation}\label{eq:n98}
\!\mathbf{a}_{\mathrm{first}}^{\mathrm{ne}}(\varphi_{p})\!=\!\left[1,e^{-j\frac{2\pi}{\lambda}(-x_{2}\sin\varphi_{p})},\cdots,e^{-j\frac{2\pi}{\lambda}(-x_{M_{1}}\sin\varphi_{p})}\right]^{\mathrm{T}}\!\!.
\end{equation}

Consequently, estimating the $P+1$ AoAs from (\ref{eq:u87}) can be
formulated as a $(P+1)$-sparse reconstruction problem, which cannot be solved using the LS due to the nonlinear relationship between the vector in (\ref{eq:n98})  and the AoAs. Here, we estimate the
AoAs by using the low-complexity orthogonal matching pursuit (OMP)
algorithm, where the dictionary matrix is constructed from the  manifold vectors derived in (\ref{eq:n98}):
\begin{align} \label{eee}
&\mathbf{C}_{\mathrm{first}}^{\mathrm{ne}} \in C^{M_{1}\times C_{\mathrm{ne}}} \nonumber\\
\!\!\!\!\!=&\left[\mathbf{a}_{\mathrm{first}}^{\mathrm{ne}}(-\frac{\pi}{2}),\mathbf{a}_{\mathrm{first}}^{\mathrm{ne}}(-\frac{\pi}{2}+\frac{\pi}{C_{\mathrm{ne}}}),\cdots,\mathbf{a}_{\mathrm{first}}^{\mathrm{ne}}(\frac{\pi}{2}-\frac{\pi}{C_{\mathrm{ne}}})\right],
\end{align}
where $C_{\mathrm{ne}}$ is the size of the dictionary and $-\frac{\pi}{2}+\frac{\pi c}{C_{\mathrm{ne}}}$
is the angle for the $(c+1)$th steering vector. The estimated AoAs
are denoted by $\{\hat{\varphi}_{p}\}_{p=1}^{P}$. 

\textsl{Distance Estimation:} For the far-end subarray, the index
$m_{\mathrm{fe}}$ is large and usually on the order of several hundreds, making the
second- and third-order terms in (\ref{eq:ji6=00005C}) significant
and thus non-negligible\footnote{For instance, when $r_{p}=10$ m, $\varphi_{p}=0$, $f_{c}=28$ GHz
and $m_{\mathrm{fe}}=400$, the second-order term is $\mathcal{O}(x_{m_{\mathrm{fe}}}^{2})=\frac{x_{m_{\mathrm{fe}}}^{2}\cos^{2}\varphi_p}{2r_{p}}=0.2333$.}. Therefore, higher-order effects must be considered, and the far-end
subarray must retain the full free-space near-field steering vector as in (\ref{eq:n9j9}),
given by
{\small
\!\!\begin{align}
\mathbf{a}^{\mathrm{fe}}(\varphi_{p},r_p)  \!=\!&\Big[\frac{r_{p}}{r_{p,m_{\mathrm{fe}}}}e^{-j\frac{2\pi}{\lambda}(r_{p,m_{\mathrm{fe}}}-r_{p})},\frac{r_{p}}{r_{p,m_{\mathrm{fe}}+1}}e^{-j\frac{2\pi}{\lambda}(r_{p,m_{\mathrm{fe}}+1}-r_{p})}, \nonumber\\ &\cdots,\frac{r_{p}}{r_{p,m_{\mathrm{fe}}+M_{2}-1}}e^{-j\frac{2\pi}{\lambda}(r_{p,m_{\mathrm{fe}}+M_{2}-1}-r_{p})}\Big]^{\mathrm{T}},\label{eq:n9j9-2}
\end{align}}
where $r_{p,m}=\sqrt{r_{p}-2r_{p}x_{m}\sin\varphi_{p}+x_{m}^{2}}$, for $m_{\mathrm{fe}} \leq m \leq m_{\mathrm{fe}}+M_2-1$.

Given the estimated AoAs $\{\hat{\varphi}_{p}\}_{p=1}^{P}$, the steering
vector in (\ref{eq:n9j9-2}) becomes a function of 
the distance $r_{p}$ only. Accordingly, we estimate $r_{p}$ using the
OMP algorithm with a distance-dependent dictionary defined as
\begin{align}
\mathbf{C}^{\mathrm{fe}}(\hat{\varphi}_{p})=&[\mathbf{a}^{\mathrm{fe}}(\hat{\varphi}_{p},d_{\mathrm{min}}),\mathbf{a}^{\mathrm{fe}}(\hat{\varphi}_{p},d_{\mathrm{min}}+\triangle d), \nonumber\\
&\cdots,\mathbf{a}^{\mathrm{fe}}(\hat{\varphi}_{p},d_{\mathrm{max}}-\triangle d)]\in C^{M_{2}\times C_{\mathrm{fe}}},
\end{align}
where $C_{\mathrm{fe}}$ is the dictionary size and $\triangle d=\frac{d_{\mathrm{max}}-d_{\mathrm{min}}}{C_{\mathrm{fe}}-1}$
denotes the distance resolution from the range $[d_{\mathrm{min}},d_{\mathrm{max}}]$. Each column $\mathbf{a}^{\mathrm{fe}}(\hat{\varphi}_{p},d_{\mathrm{min}}+c\triangle d)$
is generated from (\ref{eq:n9j9-2}) using the angle $\hat{\varphi}_{p}$
and a discrete distance value. The estimated distances are denoted
by $\{\hat{r}_{p}\}_{p=1}^{P}$.

	 \vspace{-2mm}
\subsection{Refined Estimation  of Angles and Distances}

The first-order Taylor approximation in (\ref{eq:n98}) may still
introduce non-negligible phase errors. For instance, substituting
$m=40$ in Footnote 1 yields a second-order term of
$\mathcal{O}(x_{m}^{2})=\frac{x_{m}^{2}\cos^{2}\varphi_p}{2r_{p}}=0.0023$, which results in an estimated angle of $\hat{\varphi}_p=-0.0108$ when using $\mathbf{C}_{\mathrm{first}}^{\mathrm{ne}}$ in (\ref{eee}) such that $x_{m}\sin\hat{\varphi}_p =x_{m}\sin\varphi_{p}-\frac{x_{m}^{2}\cos^{2}\varphi_{p}}{2r_{p}}$. This indicates  a maximum angle estimation error of $\varphi-\hat{\varphi}_p=0.0108$ at  $r_p=10$\;m. Since the distance estimates $\{\hat{r}_{p}\}_{p=1}^{P}$ are already available, the accuracy of angle estimation can be further enhanced
by utilizing the second-order Taylor approximation of the distance,
i.e., $r_{p,m}\approx r_{p,m}^{\mathrm{se}}=r_{p}-x_{m}\sin\varphi_{p}+\frac{x_{m}^{2}\cos^{2}\varphi_{p}}{2r_{p}}$, for $1\leq m\leq M_{1}$, which reduces the phase approximation error in vector (\ref{eq:n98}) for the near-end subarray.

Specifically, we update the angle-related dictionary denoted by $\mathbf{C}_{\mathrm{se}}^{\mathrm{ne}}$,
where each column is generated using the following steering vector
\begin{align}
\mathbf{a}_{\mathrm{se}}^{\mathrm{ne}}(\theta,\hat{r}_{p}) 
=&\Big[1,e^{-j\frac{2\pi}{\lambda}(-x_{2}\sin\theta+\frac{x_{2}^{2}\cos^{2}\theta}{2\hat{r}_{p}})}, \nonumber \\ &\cdots,e^{-j\frac{2\pi}{\lambda}(-x_{M_{1}}\sin\theta+\frac{x_{M_{1}}^{2}\cos^{2}\theta}{2\hat{r}_{p}})}\Big]^{\mathrm{T}}.\label{eq:n9j7}
\end{align}

After obtaining the refined angle estimates, denoted by $\{\hat{\varphi}_{p}^{\mathrm{se}}\}_{p=1}^{P}$,
using $\mathbf{C}_{\mathrm{se}}^{\mathrm{ne}}$ in the OMP,  the distances are re-estimated accordingly, yielding the refined distance
estimates $\{\hat{r}_{p}^{\mathrm{se}}\}_{p=1}^{P}$.

\subsection{Path Gain Estimation}

By stacking the  signals received by the near-
and far-end subarrays, the overall observation can be expressed as
\begin{align*}
\mathbf{y}_{\mathrm{n-f}} & =\left[\begin{array}{c}
\mathbf{y}_{\mathrm{ne}}\\
\mathbf{y}_{\mathrm{fe}}
\end{array}\right]\\
 & \triangleq\mathrm{diag\left(\underbrace{\left[\begin{array}{c}
\mathbf{w}_{\mathrm{ne}}\\
\mathbf{w}_{\mathrm{fe}}
\end{array}\right]}_{\mathbf{w}_{\mathrm{n-f}}}\odot\underbrace{\left[\begin{array}{c}
\mathbf{s}_{\mathrm{ne}}\\
\mathbf{s}_{\mathrm{fe}}
\end{array}\right]}_{\mathbf{s}_{\mathrm{n-f}}}\right)}\underbrace{\left[\begin{array}{c}
\mathbf{h}_{\mathrm{ne}}\\
\mathbf{h}_{\mathrm{fe}}
\end{array}\right]}_{\mathbf{h}_{\mathrm{n-f}}}+\underbrace{\left[\begin{array}{c}
\mathbf{z}_{\mathrm{ne}}\\
\mathbf{z}_{\mathrm{fe}}
\end{array}\right]}_{\mathbf{z}_{\mathrm{n-f}}}\\
 & \triangleq\mathrm{diag\left(\mathbf{w}_{\mathrm{n-f}}\odot\mathbf{s}_{\mathrm{n-f}}\right)}\mathbf{A}_{\mathrm{n-f}}{\bf b }+\mathbf{z}_{\mathrm{n-f}},
\end{align*}
where ${\bf b }=[\beta_{1},\cdots,\beta_{P}]^{\mathrm{T}}$
denotes the path gain vector, and $\mathbf{A}_{\mathrm{n-f}}=[\mathbf{a}_{1}^{\mathrm{n-f}},\cdots,\mathbf{a}_{P}^{\mathrm{n-f}}]$
is the combined steering matrix, with each $\mathbf{a}_{p}^{\mathrm{n-f}}$
containing the corresponding entries from the full steering vector
$\mathbf{a}_{p}$.

Substituting the refined angle and distance estimates $\{\hat{\varphi}_{p}^{\mathrm{se}}\}_{p=1}^{P}$
and $\{\hat{r}_{p}^{\mathrm{se}}\}_{p=1}^{P}$ into $\mathbf{A}_{\mathrm{n-f}}$
yields the estimated matrix $\hat{\mathbf{A}}_{\mathrm{n-f}}$. The
LS estimate of path gain vector $\bf b$ is then
given by
\begin{equation}
\hat{{\bf b}}=\left(\mathrm{diag\left(\mathbf{w}_{\mathrm{n-f}}\odot\mathbf{s}_{\mathrm{n-f}}\right)}\hat{\mathbf{A}}_{\mathrm{n-f}}\right)^{-1}\mathbf{y}_{\mathrm{n-f}}. \label{x3x3}
\end{equation}

\subsection{Channel Reconstruction}
Once the AoAs, distances, and channel gains are estimated, we can reconstruct $\mathbf{h}$ based on the free-space geometric model. In particular, we have
\begin{equation}\label{G_hat}
	{\hat {\mathbf{h}}} = \sum_{p=0}^{P}\hat{\beta}_{p}\mathbf{a}_{p}(\hat{r}_{p}^{\mathrm{se}},\hat{\varphi}_{p}^{\mathrm{se}}),
\end{equation}
where $\mathbf{a}_{p}(\hat{r}_{p},\hat{\varphi}_{p})$ can be calculated based on (\ref{eq:n9j9}).  The overall channel estimation algorithm is summarized in Algorithm 	\ref{Algorithm-G-1}.

\begin{algorithm}
	\caption{Proposed Estimation Based on Sparse Array}
	\label{Algorithm-G-1} \begin{algorithmic}[1] \REQUIRE ${\bf y}_{\rm ne}$, ${\bf y}_{\rm fe}$, $\mathbf{C}_{\mathrm{first}}^{\mathrm{ne}}$,  $\mathbf{C}_{{\mathrm{se}},p}^{\mathrm{ne}}$, and $\mathbf{C}_{p}^{\mathrm{fe}}$.
		
		\STATE Estimate AoAs $\{\hat{\varphi}_{p}\}_{p=1}^{P}$ using OMP with dictionary $\mathbf{C}_{\mathrm{first}}^{\mathrm{ne}}$. 
		
		\STATE 		Estimate distances  $\{\hat{r}_{p}\}_{p=1}^{P}$ by constructing dictionary $\{\mathbf{C}^{\mathrm{fe}}(\hat{\varphi}_{p})\}_{p=1}^{P}$ and applying OMP.
		
		\STATE Estimate AoAs $\{\hat{\varphi}_{p}^{\mathrm{se}}\}_{p=1}^{P}$ by  constructing dictionary $\{\mathbf{C}_{{\mathrm{se}}}^{\mathrm{ne}}(\hat{r}_{p})\}_{p=1}^{P}$ and using OMP.
		
		\STATE Estimate distances  $\{\hat{r}_{p}^{\mathrm{se}}\}_{p=1}^{P}$ by constructing dictionary $\{\mathbf{C}^{\mathrm{fe}}(\hat{\varphi}_{p}^{\mathrm{se}})\}_{p=1}^{P}$ and applying OMP.
			
		\STATE Estimate path gains $\hat{{\bf b}}$ using (\ref{x3x3}). 

		\ENSURE ${\hat {\mathbf{h}}} = \sum_{p=0}^{P}\hat{\beta}_{p}\mathbf{a}_{p}(\hat{r}_{p}^{\mathrm{se}},\hat{\varphi}_{p}^{\mathrm{se}})$.
	\end{algorithmic} 
	 \vspace{-1mm}
\end{algorithm}

	 \vspace{-4mm}
\section{Simulation Results}

In this section, we present extensive simulation results to evaluate the performance of the proposed full CSI reconstruction method. A waveguide of length 3 meters operating at a carrier frequency of 28 GHz is considered. The waveguide refractive index is $\eta_{g}=1.4$.
The user is located at a distance of 5\;m, and the distances of the $P = 2$ scatterers are randomly drawn from the range [3, 10]\;m. The angles of the multipath components are randomly drawn from the interval $[-\pi/2, \pi/2]$.  The noise power is set to $-100$ dBm, and the starting index of the far-end subarray is set to $m_{\mathrm{fe}}= 450$. Unless otherwise specified, the subarray sizes are configured as $M_1 = M_2 = 30$.

We  evaluate two metrics: 1) the normalized mean square
error (NMSE) of the full CSI, defined as $
\mathrm{NMSE} =\mathbb{E}\left\{ \frac{||\hat{\mathbf{h}}-\mathbf{h}||_{2}^{2}}{||\mathbf{h}||_{2}^{2}}\right\}$, where the expectation is taken over different channel realizations. 2) the data rate, calculated as ${\rm Rate}=\log_2(1+\frac{q_c|w_{m_{\rm opt}}h_{m_{\rm opt}}|^2}{\sigma^{2}})$, with $q_c=40\;$dBm denoting the transmit power for communcation, and  ${m_{\rm opt}}$ being the optimal pinching antenna index based on the estimated full CSI  ${\hat {\mathbf{h}}}$. We  evaluate four algorithms: Pro. Refine, which reconstructs the full CSI using refined angle and distance estimates via (\ref{G_hat}); Pro. Coarse, which estimates the path gains and reconstructs the CSI using coarse estimates $\{\hat{\varphi}_{p}\}_{p=1}^{P}$
and $\{\hat{r}_{p}\}_{p=1}^{P}$, serving as a performance lower bound; Oracle, which assumes perfect angle knowledge available at the BS while estimating the distances and path gains   in Steps 2 and 5 of Algorithm 1, respectively, thereby representing an upper bound on performance; and LS-full CSI, which directly estimates the full CSI using the LS method in (\ref{eq:u90}) with $M=2L/\lambda=560$ pinching antenns, and  serves as a benchmark.  To ensure a fair comparison, the transmit power of the pilot signal is kept constant across all schemes. Accordingly, the pilot signal is set as  $s(t)=q/M$
 for the LS method, and as  $s(t)=q/(M_1+M_2)$ for the pilot signals used in the remaining schemes.

\begin{figure}
	\centering
	\includegraphics[scale=0.4]{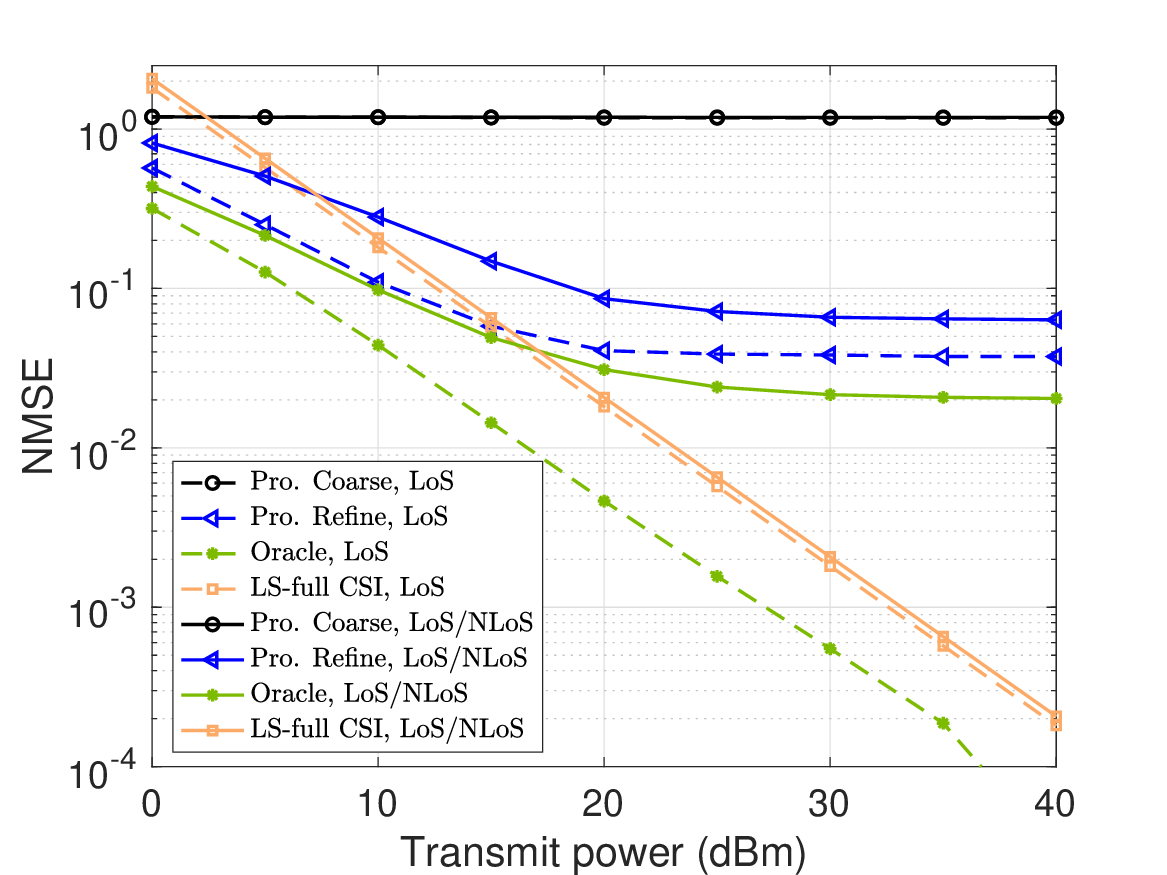}
	\vspace{-0.5em}
	\captionsetup{font={small}} 
	\caption{NMSE versus transmit power.}\label{NMSE_vs_power}
	\vspace{-1em} 
\end{figure}

Fig. \ref{NMSE_vs_power} illustrates the NMSE performance as a function of transmit power for different channel reconstruction methods. As can be observed, the proposed refined estimation approach significantly improves the accuracy compared to the coarse estimation approach. However, the refined method suffers a slight performance degradation compared to the Oracle scheme, indicating that the accuracy of angle estimation has a non-negligible impact on the subsequent distance and gain estimation. The proposed scheme outperforms the LS method for low transmit powers, while exhibiting a saturation effect at higher power levels due to the use of the OMP algorithm, whose reliance on a predefined dictionary makes it inherently vulnerable to grid mismatch, thus limiting estimation accuracy.
As expected, the LS achieves high estimation accuracy at high transmit power, but at the cost of activating a large number of antennas and significant pilot overhead.

\begin{figure}
	\centering
	\includegraphics[scale=0.4]{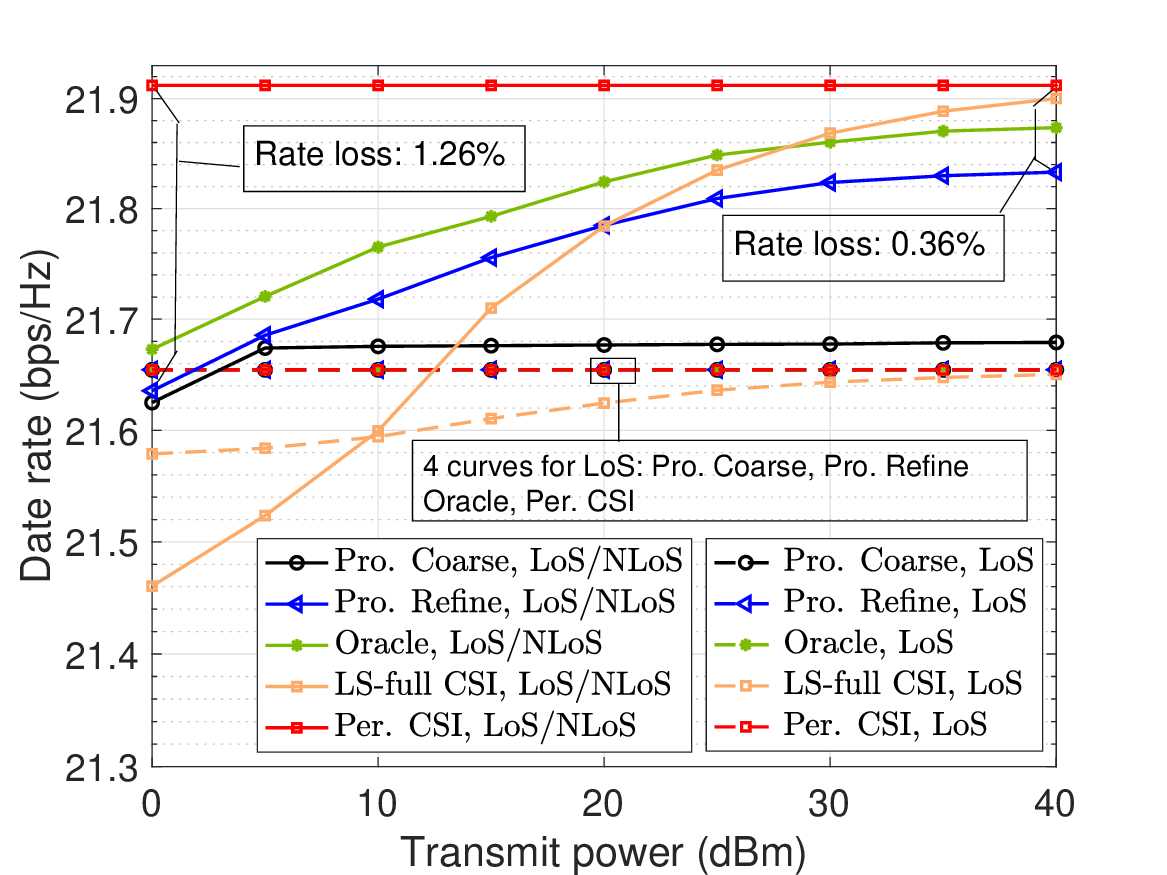}
	\vspace{-0.5em}
	\captionsetup{font={small}} 
	\caption{Data rate versus transmit power.}\label{NMSE_vs_Rician}
	\vspace{-1mm}
\end{figure}

In Fig. \ref{NMSE_vs_Rician},  the data rate corresponding to the estimated full CSI is shown as a function of the transmit power in Fig. \ref{NMSE_vs_power}. In a mixed LoS/NLoS environment, the proposed refined estimation method outperforms the LS method when the transmit power is  $\leq20$\;dBm. Although the LS method eventually outperforms the proposed approach as transmit power increases, the refined method exhibits a minimal rate loss of only 0.36\% compared to the perfect CSI case, achieving near-optimal performance. In the pure LoS scenario, the data rate from the three geometry-based estimation schemes exactly matches that under perfect CSI. This is likely because the selected index $m_{\rm opt}$ lies close to the truly optimal index $m_{\rm per}$ derived  for  perfect CSI.  As a result, the received signals at these two locations differ only in phase, not in magnitude, resulting in the same date rate. In contrast, the LS method estimates each CSI entry independently, making  $m_{\rm opt}$ more likely to deviate from the neighborhood of $m_{\rm per}$,  resulting in a larger performance gap at low transmit powers.

\begin{figure}
	\centering
	\includegraphics[scale=0.4]{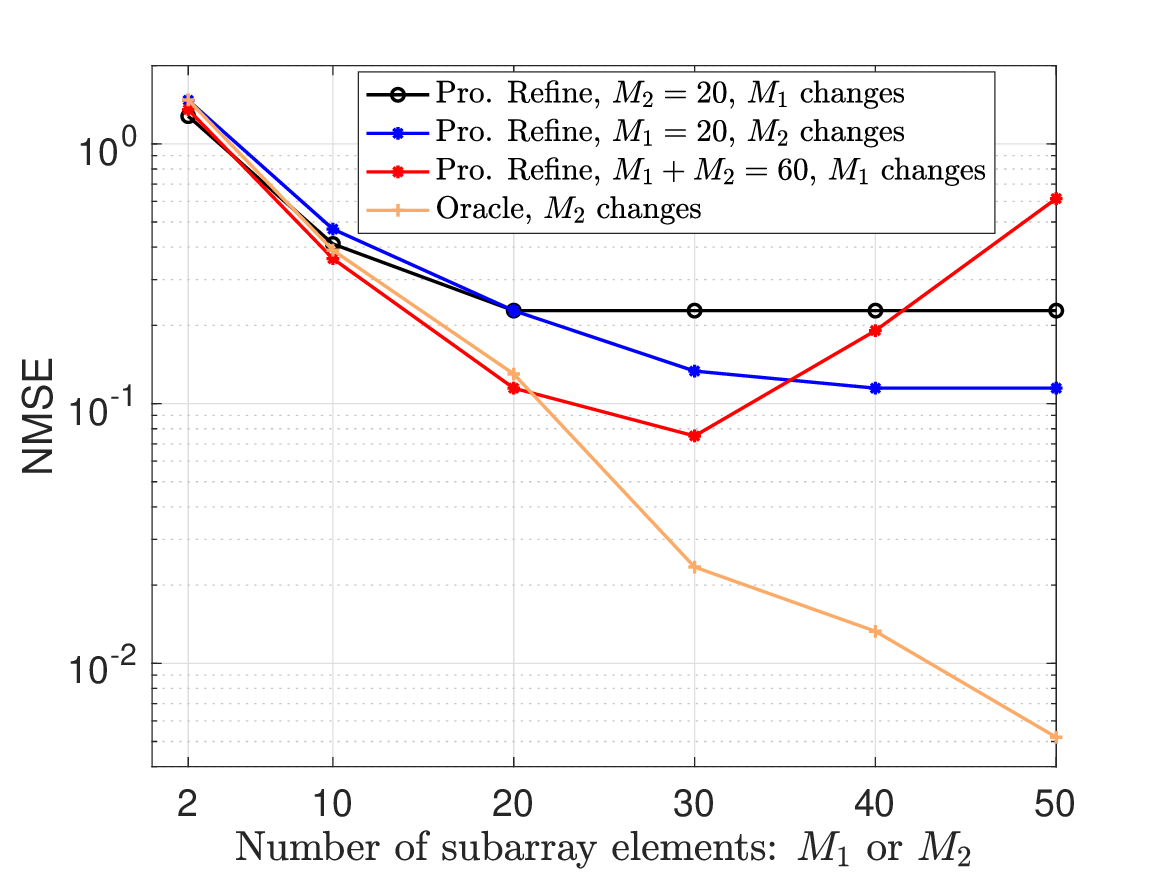}
	\vspace{-0.5em}
	\captionsetup{font={small}} 
	\caption{NMSE versus the number of pinching antennas, when $q = 40$~dBm.}\label{NMSE_vs_element}
	\vspace{-1mm}
\end{figure}

Fig. \ref{NMSE_vs_element} shows the impact of the number of activated pinching antennas in the subarrays.  For a fixed value on the $x$-axis, the total number of activated elements is identical for both the black and blue curves. As can be abserved,  configurations with  $M_1 > M_2$ consistently outperform those with  $M_1 < M_2$.  This is because  that allocating more antennas to the near-end subarray enhances angle estimation, thereby reducing the error propagated to the subsequent distance estimation. In contrast, prioritizing the far-end subarray for improved distance estimation is less effective. Furthermore, the red curve shows that, when the total number of subarray antennas is fixed, an equal allocation between the two subarrays achieves the best overall estimation accuracy.

\section{Conclusion}
In this work, we proposed a novel channel estimation scheme for mmWave PAS. A sparse large-aperture array is constructed by activating a small number of pinching antennas at both the near and far ends of the waveguide relative to the feed point. The pilot signals received by the near-end subarray are utilized to estimate the angles of arrival, while those received by the far-end subarray are used to estimate propagation distances. The path gain estimation are then  based on all the received pilot signals. Simulation results validate the effectiveness of the proposed scheme in acquiring  accurate CSI with reduced hardware complexity.

	 \vspace{-2mm}
\bibliographystyle{IEEEtran}
\bibliography{bibfile}

% Generated by IEEEtran.bst, version: 1.14 (2015/08/26)
\begin{thebibliography}{10}
\providecommand{\url}[1]{#1}
\csname url@samestyle\endcsname
\providecommand{\newblock}{\relax}
\providecommand{\bibinfo}[2]{#2}
\providecommand{\BIBentrySTDinterwordspacing}{\spaceskip=0pt\relax}
\providecommand{\BIBentryALTinterwordstretchfactor}{4}
\providecommand{\BIBentryALTinterwordspacing}{\spaceskip=\fontdimen2\font plus
\BIBentryALTinterwordstretchfactor\fontdimen3\font minus
  \fontdimen4\font\relax}
\providecommand{\BIBforeignlanguage}[2]{{%
\expandafter\ifx\csname l@#1\endcsname\relax
\typeout{** WARNING: IEEEtran.bst: No hyphenation pattern has been}%
\typeout{** loaded for the language `#1'. Using the pattern for}%
\typeout{** the default language instead.}%
\else
\language=\csname l@#1\endcsname
\fi
#2}}
\providecommand{\BIBdecl}{\relax}
\BIBdecl

\bibitem{decomo}
A.~Fukuda, H.~Yamamoto, H.~Okazaki, Y.~Suzuki, and K.~Kawai, ``Pinching
  antenna: Using a dielectric waveguide as an antenna,'' \emph{NTT DOCOMO
  Technical J.}, vol.~23, no.~3, pp. 5--12, Jan. 2022.

\bibitem{Ding-1}
Z.~Ding, R.~Schobe, and H.~V. Poor, ``Flexible-antenna systems: A
  pinching-antenna perspective,'' \emph{IEEE Trans. Commun.}, pp. 1--1, Mar.
  2025, Early Access.

\bibitem{Ding-2}
K.~Wang, Z.~Ding, and R.~Schober, ``Antenna activation for {NOMA} assisted
  pinching-antenna systems,'' \emph{IEEE Commun. Lett.}, Mar. 2025, Early
  Access.

\bibitem{Ali}
A.~Bereyhi \emph{et~al.}, ``{MIMO-PASS}: Uplink and downlink transmission via
  {MIMO} pinching-antenna systems,'' Mar. 2025.

\bibitem{FASCE}
H.~Xu \emph{et~al.}, ``Channel estimation for {FAS}-assisted multiuser {mmWave}
  systems,'' \emph{IEEE Commun. Lett.}, Mar. 2024.

\bibitem{channel-1}
S.~Lv, Y.~Liu, and Z.~Ding, ``Beam training for pinching-antenna systems
  (pass),'' Feb. 2025.

\bibitem{channel-2}
J.~Xiao, J.~Wang, and Y.~Liu, ``Channel estimation for pinching-antenna systems
  {(PASS)},'' Mar. 2025.

\bibitem{NF-channel-2}
Z.~Dong, X.~Li, and Y.~Zeng, ``Characterizing and utilizing near-field spatial
  correlation for {XL-MIMO} communication,'' \emph{IEEE Trans. Commun.},
  vol.~72, no.~12, pp. 7922--7937, Dec. 2024.

\bibitem{NF-channel}
H.~Lu \emph{et~al.}, ``A tutorial on near-field {XL-MIMO} communications toward
  {6G},'' \emph{IEEE Commun. Surv. Tut.}, vol.~26, no.~4, pp. 2213--2257,
  Fourthquarter 2024.

\bibitem{NFchannel}
M.~Cui and L.~Dai, ``Channel estimation for extremely large-scale {MIMO}:
  Far-field or near-field?'' \emph{IEEE Trans. Commun.}, vol.~70, no.~4, pp.
  2663--2677, Apr. 2022.

\bibitem{NFMUSIC}
D.~G\"{u}rg\"{u}no\u{g}lu \emph{et~al.}, ``Performance analysis of a {2D-MUSIC}
  algorithm for parametric near-field channel estimation,'' \emph{IEEE Wireless
  Commun. Lett.}, Mar. 2025, Early Access.

\end{thebibliography}

\end{document}